\documentclass[aps,prl,reprint,groupedaddress]{revtex4-1}
\usepackage{graphicx,siunitx}
\usepackage[english]{babel}
\usepackage{pdfpages}
\usepackage{etoolbox} 
\usepackage{float}
\usepackage{amsmath}

\makeatletter
\patchcmd{\@outputpage@head}{\@ifx{\LS@rot\@undefined}{}{\LS@rot}}{}{}{}
\makeatother

\begin{document}

\title{Doppler Cooled Ions in a Compact, Reconfigurable Penning Trap}

\author{Brian J. McMahon}
\email[]{brian.mcmahon@gtri.gatech.edu}
\author{Curtis Volin}
\author{Wade G. Rellergert}

\author{Brian C. Sawyer}
\affiliation{Georgia Tech Research Institute, Atlanta, Georgia 30332, USA}

\date{\today}

\begin{abstract}
We report the design and experimental demonstration of a compact, reconfigurable Penning ion trap constructed with rare-earth permanent magnets placed outside of a trap vacuum enclosure. We describe the first observation of Doppler laser cooling of ions in a permanent magnet Penning trap. We detail a method for quantifying and optimizing the trap magnetic field uniformity in situ using a thermal beam of neutral $^{40}$Ca precursor atoms. Doppler laser cooling of $^{40}$Ca$^+$ is carried out at 0.65~T, and side-view images of trapped ion fluorescence show crystalline order for both two- and three-dimensional arrays. Measured $^{40}$Ca$^+$ trap frequencies confirm the magnetic field characterization with neutral $^{40}$Ca. The compact trap described here enables a variety of cold ion experiments with low size, weight, power, and cost requirements relative to traditional electromagnet-based Penning traps.    
\end{abstract}


\maketitle


Penning traps confine ions in three dimensions (3D) with a combination of a uniform magnetic field and quadrupolar electrostatic field. Unlike in radiofrequency (RF) Paul traps, ions in Penning traps exhibit no driven micromotion and, with the use of superconducting or permanent magnets, are confined with minimal power consumption. Traditional Penning traps consist of stacked hyperbolic or cylindrical electrode structures placed within the bore of a high-field ($\gtrsim 1$~T) electromagnet operating with normal or superconducting currents. Such traps have enabled record-setting precision measurements of charged-particle masses~\cite{bradley_penning_1999} and magnetic moments~\cite{hanneke_new_2008,sturm_g_2011,smorra_parts-per-billion_2017}. Doppler laser cooling of ions in Penning traps was first demonstrated in the late 1970s~\cite{wineland_radiation-pressure_1978}, and has facilitated frequency metrology with hyperfine transitions exhibiting $>550$ s of coherent evolution~\cite{bollinger_303-mhz_1991} as well as precision measurements of hyperfine constants~\cite{shiga_diamagnetic_2011}. Recent work with laser-cooled ions in traditional Penning traps includes implementation of sub-Doppler laser cooling~\cite{goodwin_resolved-sideband_2016,stutter_sideband_2018,jordan_near_2019}, sympathetic cooling of co-trapped molecular ions~\cite{van_eijkelenborg_sympathetic_1999} and anti-matter~\cite{jelenkovic_sympathetically_2003}, spin entanglement verified by spin squeezing~\cite{bohnet_quantum_2016}, quantum simulation of Ising magnetism~\cite{britton_engineered_2012,garttner_measuring_2017}, and optical force~\cite{biercuk_ultrasensitive_2010} (ion displacement~\cite{gilmore_amplitude_2017}) detection with yN (pm) sensitivity. 

\begin{figure}
	\includegraphics[scale=0.65]{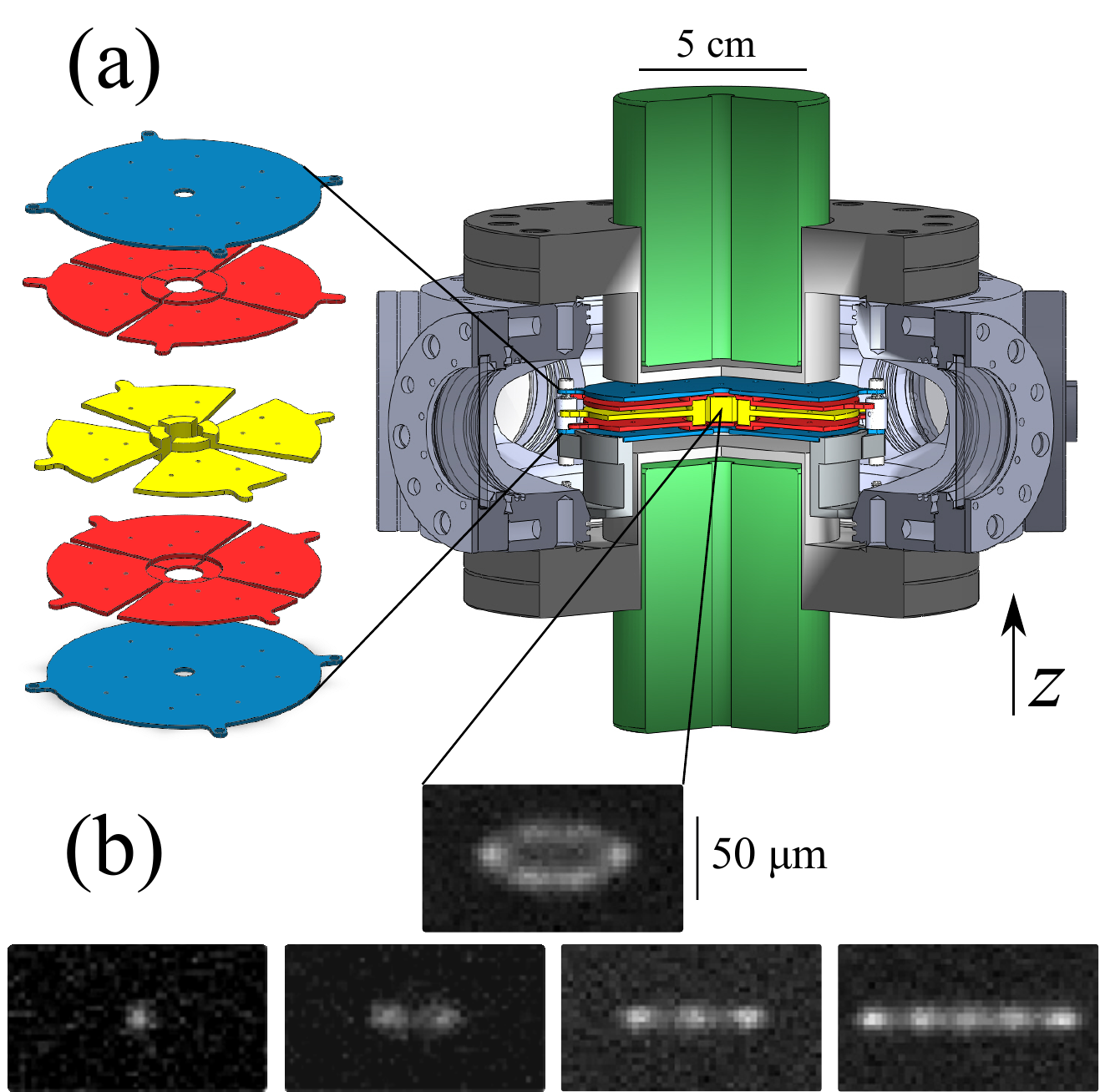}%
	\caption{\label{fig:TrapPic} (Color online) (a) Left: Axially-expanded trap electrode structure showing endcap (blue), compensation (red), and ring electrodes (yellow). Right: Section view of assembled trap. The NdFeB cylindrical ring magnets (green) are shown at the optimal spacing for maximum field uniformity. (b) Side-view images of Doppler-cooled samples of 1-20 $^{40}$Ca$^+$ ions in 3D (upper image) and 2D (bottom row) configurations.}
\end{figure}       

Miniaturization of Penning traps using permanent magnets is attractive for reduction of required laboratory infrastructure (e.g. volume, cryogens, power) and development of portable Penning traps. Replacement of large ($>1$~m$^3$) electromagnets with smaller ($\sim10^{-4}$~m$^3$) rare earth permanent magnets (REPMs) confers, for example, improved optical access for ion interrogation and imaging, shorter overall laser beam paths, and reduced fringing magnetic fields. Over the past three decades, a number of compact Penning traps based on REPMs have been demonstrated for applications including portable mass spectrometry~\cite{lemaire_compact_2018}, ion storage~\cite{gomer_compact_1995,suess_permanent_2002,tan_penning_2012}, and spectroscopy of highly-charged ions~\cite{brewer_lifetime_2018}. Rare earth magnets of the SmCo and NdFeB varieties have remanences and coercivities $> 1$~T, allowing for Penning confinement of ions with masses up to $\sim 100$~amu with a trap size, weight, power consumption, and cost each more than an order-of-magnitude below those of traditional high-field electromagnets. Notably, REPM materials have also been used in conjunction with RF Paul traps in the application of large magnetic gradients~\cite{kawai_surface-electrode_2017} and stable bias magnetic fields~\cite{ruster_long-lived_2016}. For precision measurements and quantum control experiments in Penning traps, maximizing magnetic field uniformity and stability is critical to minimizing systematic errors due to fluctuating trap and internal atomic transition frequencies~\cite{britton_vibration-induced_2016}. 

In this paper, we report the first Doppler laser cooling of ions ($^{40}$Ca$^+$) in a compact Penning trap made with REPMs. In contrast to previous work, the trap magnets reside outside the vacuum vessel, making the system quickly reconfigurable and bakeable without loss of field strength. The trap operates at room temperature and ultra-high vacuum ($<10^{-10}$~Torr), and $^{40}$Ca$^+$ samples can be trapped and cooled for days with no observed ion loss. We also document a procedure for optical characterization of the Penning trap magnetic environment using neutral precursor ($^{40}$Ca) atoms, mitigating a key technical challenge to using adjustable REPMs for cold ion experiments. This optical magnetic field characterization is applicable to a wide range of Doppler-cooled atomic ion species in both REPM- and electromagnet-based Penning traps.     

Figure~\ref{fig:TrapPic}(a) shows a drawing of the compact Penning trap used for this work. The trap electrodes are constructed as a five-layer stack of machined Ti plates (left image of Fig.~\ref{fig:TrapPic}(a)), each referenced above and below via sapphire beads and washers (not shown). On the top and bottom are two endcap electrodes (blue) with 6-mm diameter central holes for future axial laser beam access. Two sets of quarter-segmented harmonic compensation electrodes (red) allow for minimization of electric field anharmonicity and correction of misalignment between the electrode axis and trap magnetic field~\cite{heinzen_rotational_1991}. In the center of the electrode stack is the quarter-segmented ring electrode (yellow). In addition to contributing to the DC axial trap potential, the segmented electrodes permit application of a RF axialization drive~\cite{powell_improvement_2003} or a dipole/quadrupole rotating wall potential~\cite{hasegawa_stability_2005}. The compressed Ti electrode stack rests on a machinable ceramic support that is bolted inside a commercial stainless steel (316LN) octagonal vacuum chamber. Side-view imaging and laser beam access are facilitated via the four 2~mm $\times$ 6~mm gaps in the segmented ring electrode.

A pair of commercial NdFeB (specified grade N52) cylindrical ring magnets with axial magnetization (green cylinders in Fig.~\ref{fig:TrapPic}(a)) provides the vertical Penning trap magnetic field. The sintered REPMs are fixed in Al tube clamps (not shown) for position adjustments relative to the chamber. For cylindrical magnets, there exists a point along the axis where the second-order magnetic field curvature vanishes~\cite{frerichs_analytic_1992,mcmahon_supplemental_2019}. Odd-order field curvature is canceled using two axially-separated magnets. The resulting field is described to lowest order by a quartic dependence on the distance from center in both radial and axial dimensions. Each magnet used for this trap has a 6.35~cm (2.5~in) outer diameter, 0.79~cm (5/16~in) inner diameter, and is 7.62~cm (3~in) long. With this magnet geometry, we predict an optimal vertical spacing of 30 mm, a residual quartic vertical field curvature of 1.0~G/mm$^4$, and a magnetic field of $7215\ \si{\gauss} $ at the trap center. 

The thermal sensitivity of NdFeB remanence is specified to be $-0.12~\%$/K near room temperature (i.e. $-8.7$~G/K at $7215\ \si{\gauss}$), which provides a means of tuning the field magnitude. By placing the magnets outside the vacuum apparatus, we allow for realtime magnet adjustments and facilitate the use of NdFeB, which would otherwise be strongly demagnetized during our 190$^{\circ}$C vacuum chamber bakeout. Using a resistance temperature detector, we measure a magnet temperature instability of 20~mK over 1 hr. At 1~s, the temperature fluctuations are below the 5~mK instrument resolution. For durations above 1~hr, the measured magnet temperature instability matches that of the surrounding laboratory.

\begin{figure}
	\includegraphics[scale=0.5]{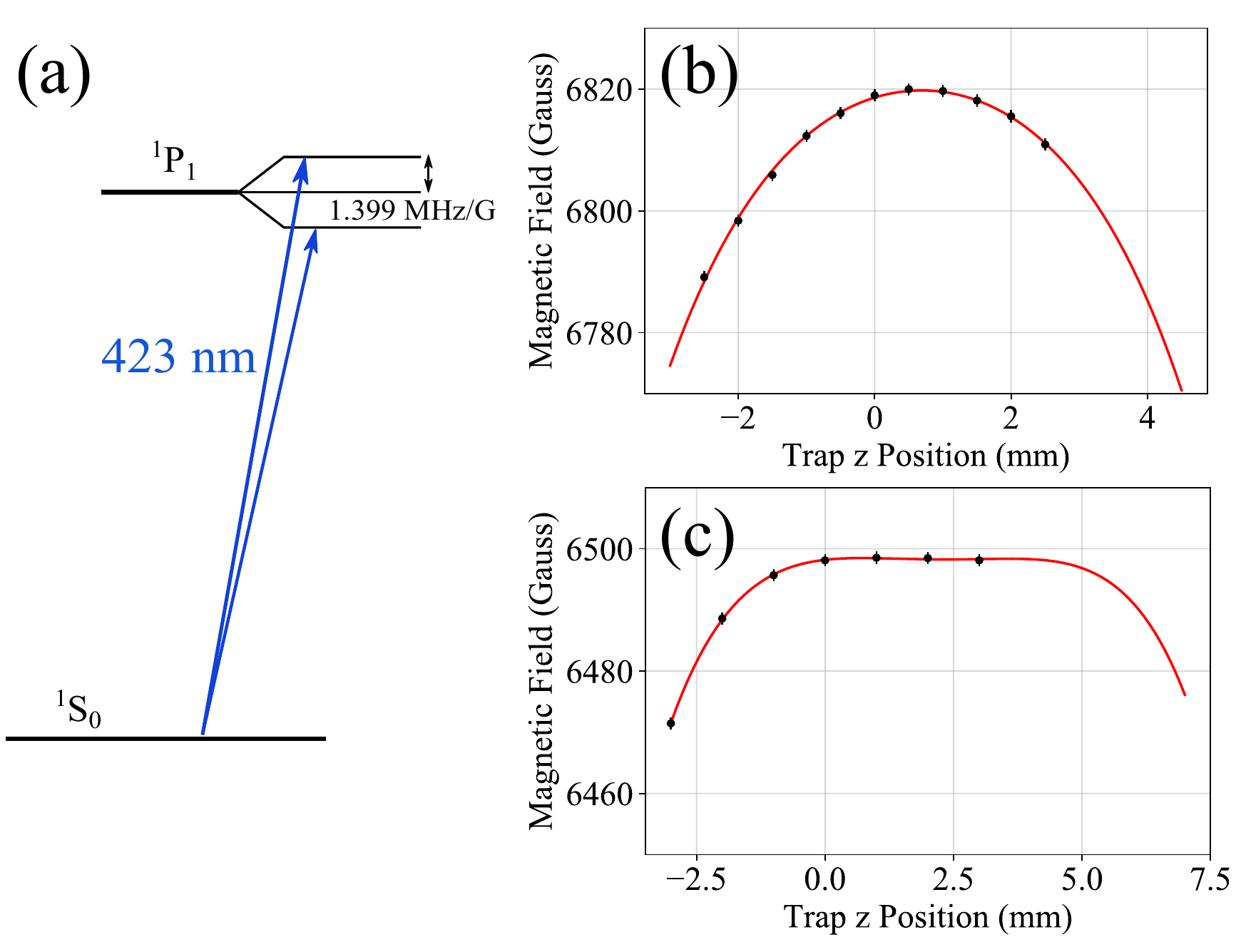}%
	\caption{\label{fig:MagneticMeas} (Color online) Measurement of vertical ($z$) trap magnetic field curvature using a thermal beam of $^{40}$Ca precursor atoms. (a) Level diagram of $^{40}$Ca showing the two transitions used for optical magnetometry. (b) Measure of the trap magnetic field as a function of axial position relative to the trap electrode center. Black points with error bars are the measured data. Red curves are fits to a multipole analytic model for cylindrical magnets~\cite{frerichs_analytic_1992}. The fit gives a magnet separation of $26.9(1)\ \si{\mm}$. (c) Same measurement as in (b) after increasing the magnet separation by 3~mm to the optimal spacing. The fit gives a final separation of 30.0(1)~mm. Error bars show standard error that includes the frequency instability of the 423 nm laser.}
\end{figure}

The trap magnetic field curvature is mapped in situ in the vertical direction using the frequency-dependent fluorescence of neutral $^{40}$Ca from an atomic oven. The atomic oven is constructed from Ta tubing and wire. The Ta tubing is filled with slivers of Ca, and the oven is mounted outside the Ti electrode structure directed radially toward the trap center. In a magnetic field, the $^1P_{1}$ state of $^{40}$Ca is split into three levels -- two magnetically-sensitive levels ($m_J=\pm1$) and one insensitive state ($m_J=0$) (see Fig.~\ref{fig:MagneticMeas}(a)). We scan a wavemeter-stabilized $423\ \si{\nm}$ laser through resonance at each of the two magnetically-sensitive $^1S_0\rightarrow ^{1}\!\! P_1$ transitions using an acousto-optic modulator (AOM). The power-stabilized 423~nm laser beam is oriented perpendicular to the nominal atomic beam velocity to avoid the large Doppler shift from the $^{40}$Ca forward velocity. The difference between the two measured optical frequencies is sensitive to the absolute magnetic field at the level of $\sim 2.799$~MHz/G. To span the large ($\sim20$~GHz) frequency difference between atomic lines, we shift the wavemeter lock point. With the wavemeter lock, we measure a RMS frequency instability below $\pm 1$~MHz as confirmed with a scanning Fabry-Perot cavity. To ensure long-term frequency accuracy, the wavemeter is calibrated to a Cs-stabilized 852~nm laser at the few-minute timescale. We map the vertical ($z$) trap magnetic field distribution by simultaneously translating the focused 423~nm beam and a fused silica imaging objective ($NA = 0.17$, $f/2.86$). The measurements are repeated at different positions within the 6~mm vertical electrode gap. Subtraction of the two optical transition frequencies at each vertical position removes the effects of frequency offsets due to slow drifts of atomic velocity, absolute 423~nm laser frequency, or laser beam angle.

Figures~\ref{fig:MagneticMeas}(b,c) show the result of our magnetic field measurements (black points with error bars) with fitted profiles (red lines). Instead of fitting the field curvature to a simple sum of polynomial terms in $z$, we fit to an analytic solution of the magnetic field due to axially-magnetized cylinders derived in Ref.~\cite{frerichs_analytic_1992}. We vary and extract three experimental parameters from the fits: the magnet spacing, vertical center offset from the electrode center, and remanence of the magnet pair~\cite{mcmahon_supplemental_2019}. The fit of the data in Fig.~\ref{fig:MagneticMeas}(b) shows that the magnet pair center is initially $0.69(2)\ \si{\mm}$ higher than our nominal trap electrode center, the magnetization is $13309(14)\ \si{\gauss}$, and the magnet spacing is smaller than intended at $26.9(1)\ \si{\mm}$. After increasing the magnet spacing and repeating the axial field measurement, the fit to the data in Fig.~\ref{fig:MagneticMeas}(c) gives a spacing $30.0(1)\ \si{\mm}$ with an offset of $2.29(5)\ \si{\mm}$ from the electrode center and a remanent magnetization of  $13395(13)\ \si{\gauss}$. The pull force between the magnets at this spacing is $>500$~N ($\sim$120~lbs.), and the magnets are strongly self-centering to a common axis. The most likely cause of the $0.6\%$ discrepancy in the remanence between the two fits is a small overall offset ($<1$~mm) of the magnet axis from the trap electrode axis. The measured magnet remanences are closer to that of grade N45 than N52 NdFeB, and the ultimate measured uniform field value is $6498(1)\ \si{\gauss}$ instead of the predicted $7215\ \si{\gauss}$.

\begin{figure}
	\includegraphics[scale=0.8]{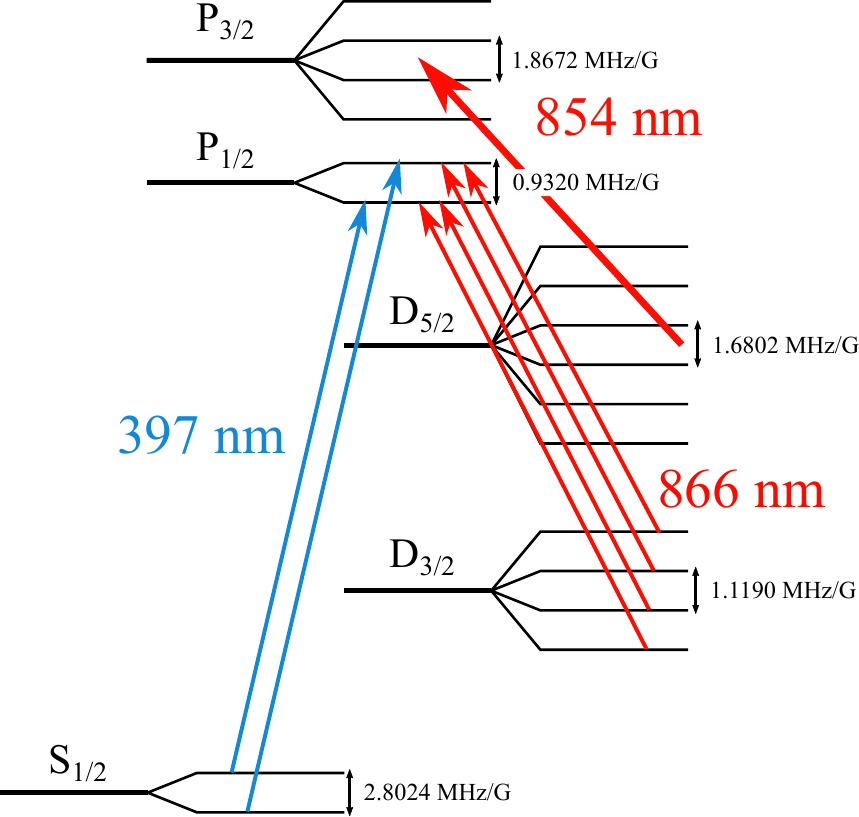}%
	\caption{\label{fig:leveldiag} (Color online) Level diagram for $^{40}$Ca$^+$, the optical transitions used in the experiment, and the magnetic sensitivity of each multiplet.}
\end{figure} 

To create $^{40}$Ca$^+$ near the trap center and ensure precise overlap of the laser beams, all photoionization (PI) and Doppler cooling laser beams are co-aligned to the aforementioned 423 nm beam and focused using a fused silica 15-cm focal length lens. For ion creation at arbitrary magnetic field, the $423\ \si{\nm}$ laser is tuned to be resonant with the field insensitive $|^1 S_{0}, m_J=0\rangle \rightarrow |^1 P_{1}, m_J=0\rangle$ transition (see Figure~\ref{fig:MagneticMeas}(a)). A second photon with a wavelength below $\sim$390 nm is sufficient for Ca ionization. We have loaded $^{40}$Ca$^+$ with $\sim1$ mW of CW 375~nm or 313~nm light as well as with a pulsed 266 nm source ($\sim10~\mu$J, 30~ps pulse duration) produced from the fourth harmonic of a 1064~nm Nd:YAG laser. We vary the number of ions loaded by adjusting the power of the second PI laser beam (see Fig.~\ref{fig:TrapPic}(b)). 

\begin{figure*}
	\includegraphics[scale=0.55]{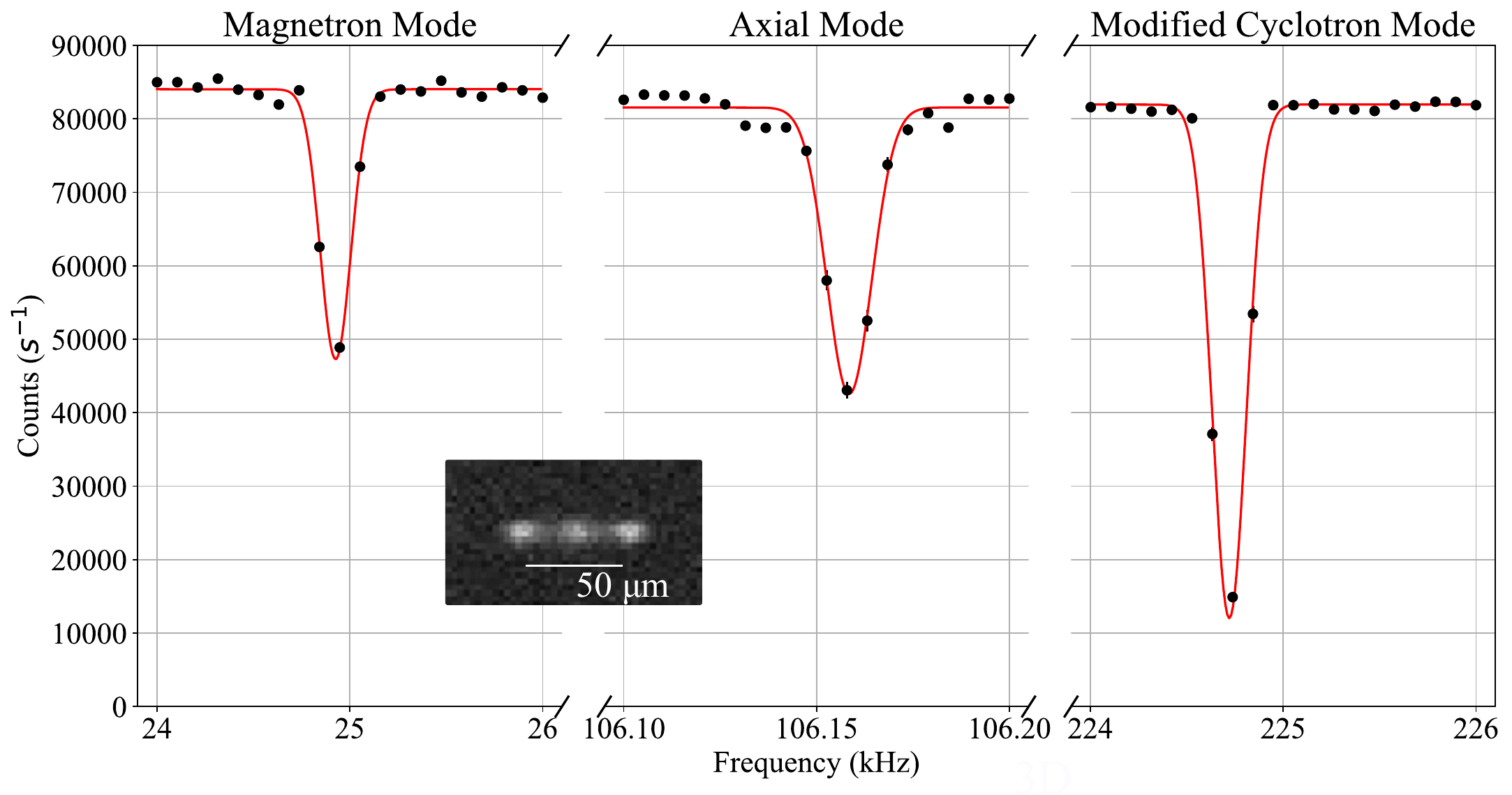}%
	\caption{\label{fig:modefreqs} (Color online) Measurement of Penning trap secular frequencies obtained from the ion fluorescence response to a weak RF drive applied to a compensation electrode. The raw data (black points with error bars) and Gaussian fits (red curves) yield the three mode frequencies (magnetron, axial, and modified cyclotron). (Inset) A side-view image of the $<10$-ion planar array used for these measurements.}
\end{figure*}

At the magnetic field values used in this work, the linear Zeeman effect splits the energy levels of $^{40}$Ca$^+$ by more than $10\ \si{\ghz}$ (see Fig.~\ref{fig:leveldiag}). For this reason, more laser frequencies are required to efficiently Doppler cool the trapped ion cloud than are needed at low field~\cite{koo_doppler_2004}. We use two separate external cavity diode lasers (ECDLs) near $397~\si{\nm}$ to drive the $|S_{1/2},m_J=\pm1/2\rangle\rightarrow|P_{1/2},m_J=\mp1/2\rangle$ transitions. One also needs four repump frequencies near $866~\si{\nm}$ to empty the long-lived $D_{3/2}$ level. Those frequencies are generated by tuning a single $866~\si{\nm}$ ECDL to the unshifted repump frequency and coupling it through a broadband fiber electro-optic modulator (EOM). Two different modulation frequencies are applied to the EOM, creating the four sideband laser tones needed for repumping out of the $D_{3/2}$ states. Further complicating the cooling process, large external magnetic fields couple like-$m_J$ states across fine structure doublets (e.g. $P_{1/2}$ and $P_{3/2}$, $D_{3/2}$ and $D_{5/2}$)~\cite{crick_magnetically_2010}. We apply an 854~nm laser beam to repump ions from the $D_{5/2}$ level back into the cooling cycle, and in practice we observe the same ion fluorescence whether or not sidebands are applied to the 854~nm light. We measure a $\sim50\%$ reduction in $^{40}$Ca$^+$ fluorescence with the 854~nm laser beam blocked. This suggests that our observed rate for dipole-forbidden $P_{1/2}\rightarrow D_{5/2}$ decays is similar to the $D_{5/2}$ spontaneous decay rate of $\sim1.2$~s$^{-1}$. 

Simultaneous application of the six 397/866 nm laser tones necessary for Doppler cooling naturally leads to the formation of dark resonances at various laser frequency combinations. To simplify initial observation of $^{40}$Ca$^+$, we start by trapping in a RF Paul trap configuration as in Ref.~\cite{koo_doppler_2004}, but here we use movable permanent magnets instead of electromagnet coils. We begin with only $\sim3\ \si{\gauss}$ bias obtained from a 25~mm diameter NdFeB ring magnet placed outside the vacuum chamber. The confining RF pseudopotential is generated by applying a 2.71 MHz sinusoidal RF drive to both endcap electrodes with a peak-to-peak RF amplitude of $500\ \si{V}$. Once initial trapping is observed and cooling is optimized at low magnetic field, we increase the field by moving the permanent magnet closer to the trap region until the ion fluorescence substantially decreases. We then re-optimize the Doppler laser cooling frequencies to maximize ion fluorescence. The magnets are replaced with larger ones to continually raise the field until we reach the final magnet configuration. At several points in this procedure, we stop to tune the 423 nm laser to a field \textit{sensitive} transition and record the resonant laser frequency of neutral $^{40}$Ca as measured by the wavemeter. By recording the resonant frequency of the neutral Ca fluorescence and the corresponding optimized 397/866~nm laser cooling frequencies, we produce a look-up table for future reference that avoids deleterious dark resonances induced by the six laser cooling tones. By characterizing the field this way, we can repeatably return to the reference points in this process by finding the magnet placement that gives maximal neutral Ca fluorescence at the recorded field-sensitive $423\ \si{\nm}$ laser frequency. Once we reach the intended Penning trap field, we turn off the RF and load the Penning trap directly after offsetting the beam position radially to supply the proper torque to keep the cloud trapped~\cite{itano_perpendicular_1988}. 

To confine $^{40}$Ca$^+$ in a harmonic axial potential, we apply -4~V to the segmented ring, +1.47~V to the compensation electrodes, and 0~V to the endcaps, where 0~V refers to the common vacuum chamber ground. We have performed ion trajectory simulations based on a finite element model of the trap electric field, and predict a trap axial frequency of 107.9 kHz. All electrodes are low-pass filtered at the vacuum feedthrough with circuits exhibiting 3~dB power attenuation at $60$~Hz. We control the collective rotation rate, and hence the aspect ratio, of the ion array by adjusting either the common radial Doppler lasers' focal position or the individual 397/866~nm cooling frequencies. As shown in the bottom row of Fig.~\ref{fig:TrapPic}(c), we observe concentric circular shells in the cross sectional fluorescence of 2D arrays as in Ref.~\cite{gilbert_shell-structure_1988}. To estimate the radial ion temperature without modifying the applied optical torque, we apply a frequency-stabilized 729 nm laser beam that is resonant with the $|S_{1/2},m_J=-1/2\rangle \rightarrow |D_{5/2},m_J=-5/2\rangle$ electric quadrupole transition in $^{40}$Ca$^+$. We tune the 729~nm laser frequency with an AOM and measure the $D_{5/2}$ state shelving probability versus quadrupole laser detuning. For 2D arrays with up to 20 ions, we measure a Gaussian full width at half maximum of $17$~MHz, corresponding to a radial Doppler temperature of $<135$~mK. We note that this is a conservative upper bound on radial temperature since the measured linewidth includes sidebands from the collective ion rotation (limited to $50$~kHz for the 2D arrays presented here).     

In addition to the neutral $^{40}$Ca magnetometry, we also measure the modified cyclotron, axial, and magnetron frequencies of trapped $^{40}$Ca$^+$ to determine the trap magnetic field. One can estimate the trap magnetic field confining a small single-species plasma using the measured trap mode frequencies and the Brown-Gabrielse invariance theorem (BGIT)~\cite{brown_precision_1982,van_dyck_number_1989,porto_series_2001}. The BGIT gives the relationship of the free-space cyclotron frequency ($f_c = eB/2\pi m$) to the frequencies of the measurable motional modes as $ f_c^2 = f_-^2 + f_+^2 + f_z^2$, where $f_-$ is the magnetron frequency, $f_+$ is the modified cyclotron frequency, $f_z$ is the axial frequency, $e$ is the electron charge, $B$ is the magnetic field, and $m$ is the $^{40}$Ca$^+$ mass. 

To measure the mode frequencies, a weak RF voltage is applied to one of the harmonic compensation electrodes. The fluorescence is measured as this RF frequency is varied. Ion motion is excited when the frequency of the drive is near resonance, which leads to an increase in Doppler broadening and a decrease in photon fluorescence~\cite{wineland_laser-fluorescence_1983}. Typical scans of this type are shown in Fig.~\ref{fig:modefreqs}. These scans were taken at the optimal magnet spacing of 30~mm with a trapped 2D ion plane of $<10$ ions as depicted in the inset image of Fig.~\ref{fig:modefreqs}. Gaussian fits to these scans give a magnetron frequency of $24.926(28)\ \si{\khz}$, an axial frequency of $106.159(3)\ \si{\khz}$, and a modified cyclotron frequency of $224.721(24)\ \si{\khz}$. The measured axial frequency agrees with the trajectory simulation prediction to within 2\%, indicating minimal stray electric field gradients from trap insulators. The BGIT yields a free-space cyclotron frequency of $249.781(2)\ \si{\khz}$ or, equivalently, a magnetic field of $6500.27(6)\ \si{\gauss}$ for the pure sample of $^{40}$Ca$^+$. The mode frequency measurement was performed at the $z=0$ position of Fig.~\ref{fig:MagneticMeas}(b). The two different magnetic field measurements (neutral vs. ion) at $z=0$ were taken on different days, and their $2(1)\ \si{\gauss}$ discrepancy is consistent with our observed day-to-day magnet temperature fluctuations of $\sim100$~mK.

In conclusion, we have demonstrated a reconfigurable compact Penning trap compatible with Doppler laser cooling of single ions and trapped-ion arrays. The device described here represents a new type of portable ion trap with a wide range of potential cold-ion physics applications including atomic/molecular ion spectroscopy, quantum simulation, and portable timekeeping. Straightforward modifications of this trap could include SmCo or radially-magnetized REPMs for improved field stability or higher uniform magnetic field, respectively. Future work will include introduction of axial laser cooling via installation of re-entrant viewports and trapping of the lower-mass ion species $^9$Be$^+$ toward development of a compact frequency reference. The short-term magnetic field instability of this trap will also be characterized via electron spin resonance techniques and compared with that of traditional Penning traps~\cite{britton_vibration-induced_2016}.

The authors thank J. J. Bollinger, N. D. Guise, K. R. Brown, C. M. Seck, and R. C. Brown for helpful discussions and comments on the manuscript. This work was funded by Office of Naval Research Grant No. N00014-17-1-2408.


%

\clearpage
\section{Supplemental Material}
	
Here we describe in more detail the procedure for fitting the measured axial magnetic field curvature within the rare earth permanent magnet (REPM) Penning trap. The primary purpose of this Supplement is to provide the theoretical basis of our magnetic field characterization and, for completeness, we also examine the consequences of using magnets with different remanent magnetizations. We will assume identical geometries (radii and lengths) for the magnets, because factory dimensional tolerances of $\sim 25$~$\mu$m will not dominate observed deviations from ideal behavior for cm-scale magnets.  

Reference~\cite{frerichs_analytic_1992} provides an analytic multipole expansion of the magnetic potential created by cylindrical permanent magnets. As the derivation assumes a scalar instead of a vector magnetic potential, it is valid only in free space. The magnetic scalar potential ($\Phi_M$) for an azimuthally symmetric magnet may be expressed in the near-field limit as
\begin{equation}\label{phiM}
\Phi_M = B_r \sum_{l=1}^\infty C_l r^l P_l\left(\frac{z}{r}\right)
\end{equation}  
where $B_r$ is the remanent magnetization, $l$ is the multipole order, $C_l$ are the multipole coefficients for the given magnet geometry, $r=\sqrt{\rho^2+z^2}$ is the position coordinate, and $P_l$ is the Legendre polynomial of order $l$. Specializing to the case of axial potentials ($\rho=0$), Eq.~\ref{phiM} may be simplified to
\begin{equation}\label{phiM_sim}
\Phi_M = B_r \sum_{l=1}^\infty C_l z^l
\end{equation} 
where we have used the identity $P_l(1)=1$. The axial magnetic field at position $z$, $B_z(z)$, is then just
\begin{equation}
B_z(z)=\frac{\partial}{\partial z} \Phi_M = B_r \sum_{l=1}^{\infty} l C_l z^{l-1}.
\end{equation}

Reproducing from~\cite{frerichs_analytic_1992} the multipole coefficients for axially semi-infinite rings (extending from $z_0$ to $\infty$) with an inner (outer) radius of $\rho_1$ ($\rho_2$) and \textit{axial} magnetization, we have
\begin{equation}\label{Cl}
C_l(\rho_1,\rho_2,z_0) = \left\{ \begin{matrix}
\left. \frac{1}{2} \frac{z_0}{\sqrt{z_0^2 + \rho^2}} \right \rvert_{\rho_1}^{\rho_2}, & l=1 \\ \\
\left. \frac{1}{2l(l-1)} \frac{\rho P_{l-1}^1\left(z_0/\sqrt{\rho^2+z_0^2}\right)}{(\rho^2+z_0^2)^{l/2}} \right \rvert_{\rho_1}^{\rho_2}, & l>1
\end{matrix}
\right . .
\end{equation}  
To represent a magnet of finite length, $L$, spanning $[z_0,z_0+L]$ with an overall magnetization in the $+z$ direction, we define
\begin{equation}\label{Clz}
\widetilde{C}_l(\rho_1,\rho_2,z_0,L) \equiv C_l(\rho_1,\rho_2,z_0) - C_l(\rho_1,\rho_2,z_0+L) .
\end{equation}

\subsection{Identical Magnetizations, Fit Function}

Following Eq.~\ref{Clz}, the total magnetic field, $B_z^{tot}(z)$, due to a pair of identical magnets spaced apart by a vertical gap, $g$, and an overall offset of $\Delta z$ from the origin is
\begin{equation}\label{fit}
\begin{split}
B_z^{tot}(z) &= B_r \sum_{l=1}^{\infty} \left[ \widetilde{C}_l(\rho_1,\rho_2,g/2,L) + \widetilde{C}_l(\rho_1,\rho_2,-g/2,L) \right ] \\
& \quad \times l (z-\Delta z)^{(l-1)}.
\end{split}
\end{equation}
The symmetry of the individual multipole coefficients $\left( C_l(\rho_1,\rho_2,-z_0) = (-1)^l C_l(\rho_1,\rho_2,z_0) \right)$ for axially-magnetized rings~\cite{frerichs_analytic_1992} allows us to further simplify $B_z^{tot}(z)$ as
\begin{equation}\label{fit_simple}
B_z^{tot}(z) = B_r \sum_{l=1}^{\infty} (1-(-1)^l) \widetilde{C}_l(\rho_1,\rho_2,g/2,L) l (z-\Delta z)^{(l-1)}.
\end{equation}

Note that only odd-order multipoles (even-order magnetic field curvatures) are non-zero for identical ring magnets in Eq.~\ref{fit_simple}. Equation~\ref{fit_simple} is used to fit the measured magnetic field profiles given in Fig. 2 of the manuscript by varying $\Delta z$, $g$, and $B_r$. As is pointed out in~\cite{frerichs_analytic_1992}, there exists an `optimal' spacing, $g_{opt}$, where the multipole coefficient $\widetilde{C}_3(\rho_1,\rho_2,z_0,L)$ from each magnet is zero at the center of the magnet gap and the leading magnetic field curvature is quartic instead of quadratic. This optimal spacing depends only on the magnet dimensions $(\rho_1,\rho_2,L)$ and is computed numerically from Eq.~\ref{Clz} in order to design a uniform-field Penning trap. We note that solid magnet cylinders ($\rho_1=0$) also produce points of vanishing $\widetilde{C}_3(0,\rho_2,z_0,L)$, but designs making use of solid cylinders do not allow axial optical access and generally have smaller $g_{opt}$ values for the same $B_z$ at trap center. 

\subsection{Different Magnetizations}

We now estimate the effect of a mismatch between the top and bottom ring magnetizations caused by, for example, a temperature difference. We will define the following average ($\bar{B}_r$) and differential ($\Delta B_r$) magnetizations as
\begin{equation}
\begin{split}
\bar{B}_r & \equiv \frac{B_r^{(t)} + B_r^{(b)}}{2} \\
\Delta B_r & \equiv \frac{B_r^{(t)} - B_r^{(b)}}{2},
\end{split}
\end{equation}
where $B_r^{(t)} (B_r^{(b)})$ refers to the top (bottom) magnet remanence. To simplify notation from this point onward, we will also assume without loss of generality that $\Delta z = 0$. With these modifications, Eq.~\ref{fit} becomes
\begin{equation}
\begin{split}
B_z^{tot}(z) &= \bar{B}_r \sum_{l=1}^{\infty} (1-(-1)^l) \widetilde{C}_l(\rho_1,\rho_2,g/2,L) l z^{(l-1)} \\
& \quad + \Delta B_r \sum_{l=1}^{\infty} (1+(-1)^l) \widetilde{C}_l(\rho_1,\rho_2,g/2,L) l z^{(l-1)}.
\end{split}
\end{equation}
We now compute the axial magnetic field curvature ($k^{th}$ derivative) at the trap center due to each multipole order, distinguishing between even and odd multipoles:
\begin{widetext}
	\begin{equation}\label{curvature}
	\left. \frac{\partial ^k B_z^{tot}(z)}{\partial z^k} \right\rvert_{z=0} = \\
	\left\{
	\begin{matrix}
	2 \bar{B}_r (k+1)! \widetilde{C}_{k+1}(\rho_1,\rho_2,g/2,L), & k \ \text{even} \\ \\
	2 \Delta B_r (k+1)! \widetilde{C}_{k+1}(\rho_1,\rho_2,g/2,L), & k \ \text{odd}
	\end{matrix} \right. .
	\end{equation}
\end{widetext}
Equation~\ref{curvature} provides two key insights for Penning trap design with REPMs: (1) the optimal magnet gap for minimizing quadratic ($k=2$) curvature is \textbf{independent} of the relative magnetizations and (2) an imbalance in magnet remanence contributes odd-order field curvature at the trap center that is otherwise absent by symmetry.

\end{document}